\documentclass[aps,preprint,amssymb]{revtex4}

\usepackage{epsfig}

\begin{document}
\renewcommand{\thefootnote}{\fnsymbol{footnote}} 
\renewcommand{\theequation}{\arabic{section}.\arabic{equation}}

\title{Diffusion in a model metallic glass: heterogeneity and ageing}

\author{H. R. Schober}
\email[]{E-mail: H.Schober@FZ-Juelich.de}

\affiliation{Institut f\"ur Festk\"orperforschung, Forschungszentrum
J\"ulich, D-52425 J\"ulich, Germany}

\date{\today}

\begin{abstract}
\noindent 
We report results of molecular dynamics simulations of a binary
Lennard-Jones system at zero pressure in the undercooled liquid
and glassy states. We first follow the evolution of diffusivity
and dynamic heterogeneity with temperature and show their
correlation. In a second step we follow the ageing of a quenched
glass. As diffusivity decreases with ageing, heterogeneity increases.
We conclude that the heterogeneity is a property of the inherent
diffusion of the relaxed state. The variations with aging time
can be explained by annealing of quenched defect structures. This
annealing has the same decay constants for both diffusivity and
heterogeneity of both components.
\end{abstract}

\maketitle

\section{Introduction}
\label{intro}
Diffusion  in glasses and their melts is fundamentally different
from diffusion in crystalline lattices. It has been studied  
in experiment intensively for many years. The increase in computer
power in recent years now makes computer simulation studies near 
and slightly below the glass transition temperature possible. 
Combining the results from experiment and simulation one  
increasingly gains insight into the underlying atomic dynamics.
Despite this effort, many questions remain and no general agreement 
on the
nature of diffusion and its change at 
temperatures near
the glass transition has been reached, even for simple densely packed 
glasses, such as binary metallic glasses. However, the research of 
recent years
has greatly improved our understanding and several major issues have been
settled; see the recent review \cite{RMP}.  

In a hot liquid, diffusion is by flow, whereas, in 
glass, well below the transition temperature, it will be mediated 
by hopping
processes. Key question are the transition between the
two regimes and also
the nature of the hopping process. Are the
jumps governed 
by a vacancy mechanism, similar to diffusion in the 
crystalline state, or by a mechanism inherent to the disordered 
structure? 

Glasses are thermodynamically not in  equilibrium; their
properties depend on the production history and
one observes ageing. The diffusion coefficient of a glass which 
has been
relaxed for a long time will be considerably lower than the one of an
``as quenched'' glass. Experimental evidence suggests that this
could involve a change of diffusion mechanism.

In the glassy state, diffusion can be described by an Arrhenius law.
The activation enthalpies are typically 1 to 3~eV \cite{RMP}. Other than
in crystals the pre-exponential factor varies widely  from
about $10^{-15}$ to $10^{+13}$~m$^2$s$^{-1}$ \cite{RMP}. 
In the undercooled liquid, diffusion is often described by a 
Vogel-Fulcher-Tammann law \cite{fulcher:25,tammann:26}
\begin{equation}
D(T) = D_0^{\rm VFT} \exp{-H/k(T-T^{\rm VFT})}
\label{eq_DVFT}
\end{equation}
which describes a stronger than exponential decrease of diffusivity
upon cooling toward $T^{\rm VFT} < T_{\rm g}$
where $T_{\rm g}$ is the glass transition temperature. 
As $T \to T^{\rm VFT}$,
a residual hopping diffusion eventually becomes essential and the
diffusion crosses over to the Arrhenius law.
Mode coupling theory gives a different expression for diffusion in 
the undercooled liquid \cite{gotze:92} 
\begin{equation}
D(T) = D_0^{\rm MCT} \left(T-T_c\right)^\gamma.
\label{eq_DMCT}
\end{equation}
The critical temperature $T_c$, where viscous flow is arrested
according to this expression, 
lies between the glass transition and melting temperatures,
$T^{\rm VFT} < T_g < T_c < T_m$. Again, hopping terms
not included in the simplified expression will become dominant
and prevent this freezing.
The diffusion data alone can be fitted by both expressions.
Neither
expression allows a clear insight into the atomic process behind
diffusion. 

From the pressure dependence of diffusion one finds, in general, 
apparent activation volumes much smaller than an atomic 
volume \cite{RMP}.
For vacancy diffusion  the activation volumes are of the order of
the formation volume, i.e. of the order of the atomic volume. 
Low activation volumes hint at diffusion without formation of
defects, i.e. diffusion by an inherent mechanism.
A key  to the nature of diffusion was found in measurements of the
isotope effect of diffusion $E$. It is defined by the ratio of the
diffusivities, $D_\alpha$, $D_\beta$ of two isotopes with masses
$m_\alpha$ and $m_\beta$, respectively $
E = \left( D_\alpha /D_\beta -1\right)/\left(\sqrt{m_\alpha /m_\beta}
   - 1\right)$
\cite{schoen:58}.
Because of the $1/\sqrt{m}$-dependence of the
atomic velocities
$E$ is of order unity for diffusion via single vacancy 
jumps in 
densely packed lattices where essentially single atoms jump
\cite{mehrer:90}. In contrast both in glasses
\cite{faupel:90} and in undercooled metallic melts \cite{ehmler:98}
nearly vanishing isotope effects were observed.
The vanishing isotope effect has a natural explanation if one 
assumes a collective diffusion mechanism both above and below the
glass transition. The usage of the term collective follows the general usage
in the field, see e.g. \cite{RMP}. It indicates that groups of
atoms jump together as opposed to the usual jumps into
vacancies in lattice diffusion.

The vanishing isotope effect is a property of relaxed 
glasses. Upon quench, defects can be frozen in. These 
enhance diffusion and can lead to an increased isotope effect which
vanishes upon relaxation \cite{ratzke:92}. Quasi vacancies can be
produced also under irradiation which again enhances diffusion, see
\cite{frank:96}.

In molecular dynamics simulations one can
follow the motion of
atoms over periods of several ns and up to $\mu$s.
Early simulations were restricted to simple model systems, 
such as Lennard-Jones \cite{wahnstrom:91} or soft sphere systems 
\cite{miyagawa:88}  above the glass transition. Now it is
possible to simulate real systems such as e.g. NiZr \cite{teichler:97} or
CuZr \cite{KS:01}. As far as qualitative properties of metallic glasses
are concerned the results of the model systems and the real ones are in
full agreement and can be used interchangeably.
 
Extensive studies of different properties of an undercooled binary 
Lennard-Jones 
system (BLS) at constant volume \cite{kob:95a,kob:95b} showed good 
agreement with the predictions of 
MCT. For the pressure
derivative this theory predicts a singularity at $T_c$
which should be modified by hopping processes to a 
sharp cusp of the apparent
activation volume at $T_c$. This was observed in a simulation of a
BLS at zero pressure \cite{S:02}. Both above the cusp,
in the undercooled liquid, and below, in the glass, 
activation volumes 
of around 0.4 atomic volumes were found. In the hot liquid the 
activation
volume rises to values near the atomic volume, which is expected
for binary collisions.

The change of the diffusional isotope effect upon cooling was studied
at zero pressure for monatomic Lennard-Jones liquids \cite{KS:00} 
and for BLS \cite{S:01}.
In a hot liquid one has values $E\approx 1$. Upon cooling, $E$ drops
to below $E=0.1$ already well above both $T_g$ and $T_c$. 
This drop is mainly
driven by densification but at different rates for the two 
components of the BLS. 

The small isotope effect can be understood from the elementary step
of diffusion, the atomic hopping.
In a soft sphere glass, at low temperatures, this atomic hopping was
found to be highly collective. Chains of ten atoms and more
move together in a jump. Whereas the total jump length, summed over all
participating atoms, is of the order of a nearest neighbour distance,
a single atom only moves a fraction thereof \cite{SOL:93,OS:99}. 
The same chain like motion was observed in snap shots of the motion
in the undercooled melt \cite{SGO:97,SGO:97b,donati:98}. This 
collective
motion is similar to the one of a group of people threading 
their way through
a crowd. They will fill whatever space they find in front -- small
activation volume, and each member will follow instantaneously the one
in front -- collectivity. Obviously they will get on better as a file
than in a broad front. Chain like motion was observed already in
early simulations of melting in two dimensions \cite{alder:62}.

Subsequent
atomic jumps are strongly correlated with each other, not only in the
case of return jumps \cite{OS:99}, and atoms which have 
just jumped have a
strongly enhanced probability to jump again, leading to local bursts
of activity interspersed with times of relative calm 
\cite{OS:99,teichler:01}.
These correlations, which are typical of all glasses, 
not only metallic ones, 
are closely related to the so called ``dynamic
heterogeneity'' of glasses and undercooled liquids. At any given time only
a few atoms are active. The resulting mean square displacements strongly
deviate from a Gaussian distribution for long times (non-Gaussianity)
before it is finally restored by long range diffusion. 

In this paper we will present results for the temperature dependence
of the dynamic heterogeneity and then study effects of ageing on both
the diffusivity and the dynamic heterogeneity.

\section{Simulation details}
The calculations are done for a BLJ system
\begin{equation}
V_{ij}(R) = 4 \epsilon_{ij} \left[ \left( \sigma_{ij} /R \right)^{12} - 
                         \left( \sigma_{ij} /R \right)^{6} +
                          A_{ij} R + B_{ij} \right] .
\end{equation}
where the subscripts $ij$ denote the two species.
The potential cutoff was set at $R_c = 3\sigma$. For the parameters we
took the commonly used values of Kob and Andersen \cite{kob:95a}:
$\epsilon_{AA}= \epsilon = \sigma_{AA} =\sigma = 1$, 
$\epsilon_{BB}=0.5$, $\sigma_{BB}=0.88$,
$\epsilon_{AB}=1.5$ and $\sigma_{AB}=0.8$.  
To avoid spurious cutoff effects we introduce the parameters
$A_{ij}$ and $B_{ij}$
to ensure continuity of the potential and its first derivative at the
cutoff, similar to 
the shifted force potential \cite{nicolas:79}. 
All masses were set to $m_j=1$.
As usual, in the following,  we will give all results in the 
reduced
units of energy $\epsilon_{AA}$, length $\sigma_{AA}$, and atomic mass 
$m_A$. To compare with real metallic glasses one can equate one
time unit ($(\epsilon/m\sigma^2)^{-1/2}$) roughly to 1~ps. 

The calculations were done with zero pressure and periodic boundary 
conditions. The time step was 
$\Delta t = 0.005$. Control runs
with $\Delta t = 0.0005$
showed no significant deviation. A heat bath was simulated 
by comparing
the temperature averaged over  20 time steps with the nominal temperature.
At each  time step 1\% of the temperature difference was adjusted by
random additions to the particle velocities. Apart from the very first
steps of the ageing procedure the correction, after excursions
of the temperature due to relaxations, did not exceed $10^{-4}$ of the
average velocity. 
This procedure assured that
existing correlations between the motion of atoms were only minimally
affected. The results did not change when the temperature adjustment
was varied within reasonable limits. 

The ageing studies were done for samples quenched to 
$T=0.32~\epsilon/k$, below the glass transition 
temperature, defined from the kink in the volume versus temperature
curve. Ageing leads densification. However for the parameters
adopted the aged system was still sufficiently far from the density
of the undercooled liquid to be considered glassy.
We proceeded from the samples prepared in our previous work on the
pressure dependence of the diffusivity  
where we had prepared  16 samples for each temperature, 8 with 
slightly 
positive and 8 with slightly negative pressure \cite{S:02}.
Each sample consists  of 5488 atoms in a 
ratio $4:1$ of $A$- and $B$-atoms. The samples were then additionally aged
for up to 32$\cdot 10^6$ time steps. Measurements were done during runs
with constant volume. 
In all runs pressure and energy were monitored to ensure stability of the
configurations. The measured pressure was used to interpolate to zero 
pressure. The diffusion constant was calculated from the asymptotic
slope of the atomic mean square displacements.

\begin{figure}[ht]
  \begin{center}
 \includegraphics*[bb=30 100 630 530,totalheight=7cm,keepaspectratio]{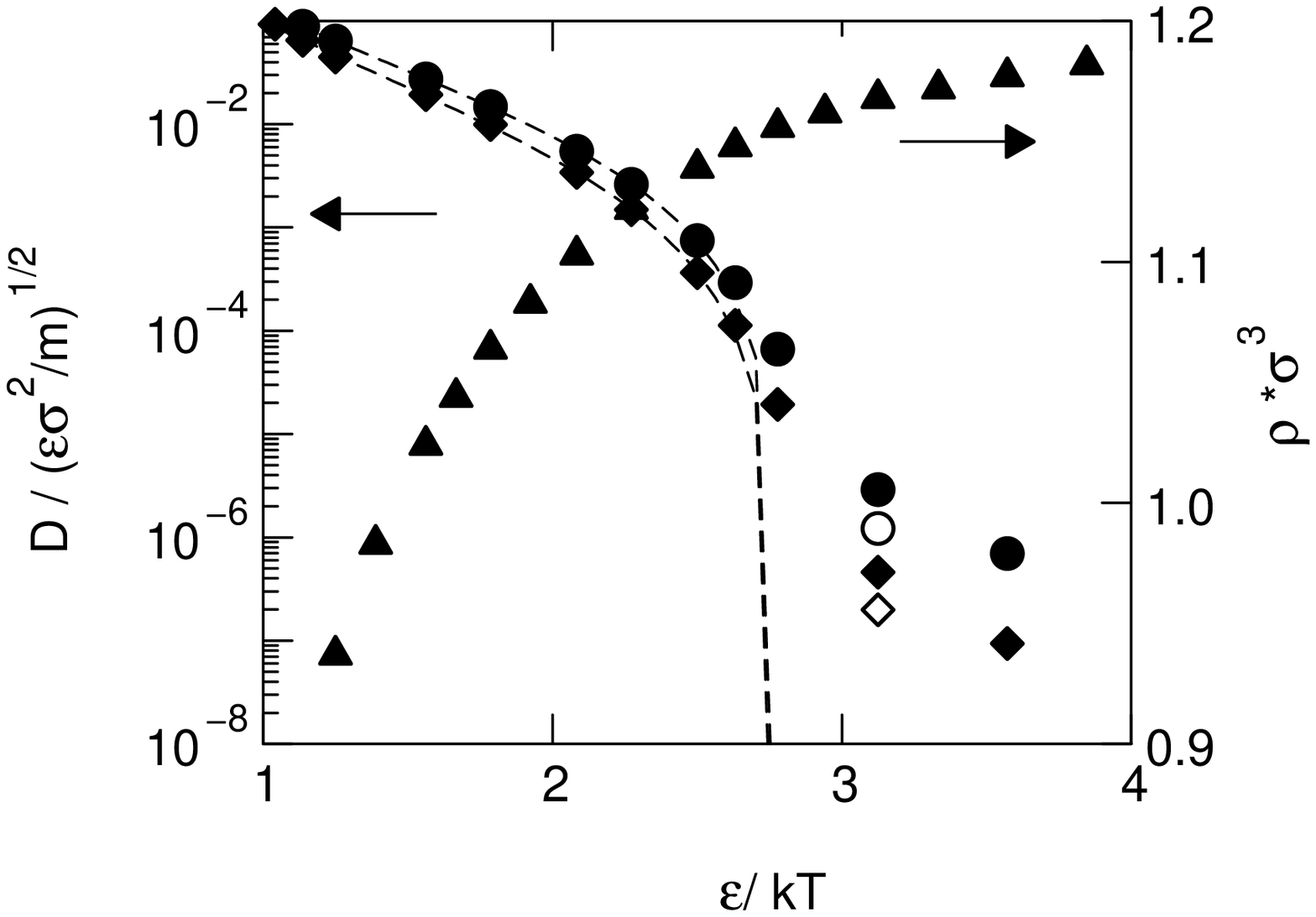}
\caption{Diffusion constants (majority $A$-atoms, diamonds, and minority $B$-atoms, 
circles) and density (triangles)
at zero pressure against inverse temperature (all in reduced units).
The open symbols indicate the extrapolated inherent diffusion
coefficients according to Eq.~\ref{eq_D_ageing}
The dashed 
and dash-dotted lines
show the fit with the mode coupling expression for atoms $A$ and $B$,respectively.}
\label{fig_D_rho} 
  \end{center}
\end{figure}

Fig.~\ref{fig_D_rho} shows the densities and diffusion constants for 
zero pressure. From the change in slope of the 
volume expansion
we estimate the glass transition temperature as $T_g \approx 0.35 \epsilon/k$.
The diffusion constant can be fitted very well with the mode coupling 
expression,
Eq.~\ref{eq_DMCT}, using a value $T_c = 0.36 \epsilon /k$ for both species and 
$\gamma = 1.87$ and $\gamma = 2.02$, respectively. The two temperatures are
very close to each other, $T_g \approx T_c$, but are much lower than the 
value $T_c = 0.435 \epsilon /k$, reported for simulations at constant 
density $\rho = 1.2$  
\cite{kob:95a}. This reflects the strong dependence of the glass 
transition
on density or pressure. We find for zero pressure a density of $\rho = 1.16$ at
$T_g$. 

The diffusion coefficients depend in a rather intricate way
both on temperature and atomic density. This makes
a comparison of the present zero pressure values
with the the previous isochoric ones \cite{kob:95a}
difficult. First the BLS becomes at $p=0$ unstable for 
$T \approx 1\epsilon/k$ whereas the high density used in \cite{kob:95a}
stabilizes the BLS to $T > 5 1\epsilon/k$. As mentioned in the
introduction the pressure derivative of $D$ (activation volume)
has a cusp at $T_c$ this implies that one would have to scale 
temperature as $T/T_c$ and subsequently scale with pressure which
again is complicated by the strong temperature variation of the
activation volume which reflects the transition from a thin liquid
dominated by binary collisions through the viscous undercooled
liquid to the solid like glass \cite{S:02}. Furthermore
it has been shown that both components are affected differently
by density. Density is a strong driving force towards cooperative
motion. However, in a binary liquid there is no longer a 
simple proportionality \cite{S:01}. We have, therefore, not attempted
to scale our values over the whole temperature range to the
ones of \cite{kob:95a}. Doing a rough scaling just above $T_c$ we
find qualitative agreement. 

Notabene,
the present values for the diffusivity below $T_{\rm g}$,
in the glass, are somewhat lower
than the ones published previously \cite{S:02} which is an effect of the
additional ageing as we will discuss further down.

\section{Dynamic heterogeneity}
\label{sect_het}
In isotropic diffusion the atomic displacements are Gaussian
distributed. In undercooled liquids and in glasses Gaussianity
is violated over long time scales. This non-Gaussianity
indicates different mobilities of different atoms over long 
time scales, the so called dynamic heterogeneity. This effect
is quantified by the 
non-Gaussianity parameter \cite{rahman:64}
\begin{equation}
\label{eq_ngp}
\alpha_2(t)=\frac{3 <\Delta r^4(t)>}{5 <\Delta r^2(t)>^2}-1,
\label{eq_alpha}
\end{equation}
where $<...>$ denotes time averaging, $\Delta r^2(t)$ is the mean
square
displacement and $\Delta r^4(t)$ the mean quartic displacement.
This parameter is defined  so that it is equal to zero when the atomic
motion is homogeneous. Experimentally it can be obtained from the
$q$-dependence of the Debye-Waller factor \cite{zorn:97}. It has 
been calculated in numerous molecular dynamics simulations
of liquids, e. g.  \cite{miyagawa:88,kob:95a,doliwa:98,CS:00,CMS:00,vollmayr:02}. 
There are three time regimes of $\alpha_2(t)$. First
it increases from $\alpha_2(t=0) = 0$ on a ps time scale due to
heterogeneities of the atomic vibrations. The maximal 
value of $\alpha_2(t)$ in this vibrational regime is small,
less than 0.2. Depending on temperature
this is followed by a strong increase during the time of the
so called $\beta$-relaxation.
At times, corresponding to the
$\alpha$-relaxation time, $\alpha_2(t)$ reaches a maximum and
drops for $t \to \infty$ to $\alpha_2(t=\infty) = 0$.
The strong increase seems to follow a $\sqrt{t}$-law, independent
of the material. This $\sqrt{t}$-law has been shown to be
a direct consequence of the collectivity of the diffusional jumps
and the correlation between subsequent jumps \cite{CMS:00}.
This general behavior holds both below and above $T_g$.  

\begin{figure}[ht]
  \begin{center}
 \includegraphics*[bb=30 100 630 530,totalheight=7cm,keepaspectratio]{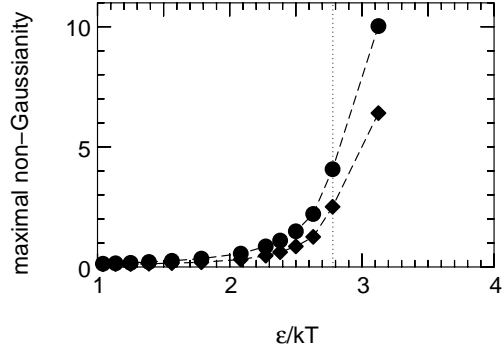}
\caption{Maximum value of the non-Gaussianity  
(majority $A$-atoms, diamonds, and minority $B$-atoms, 
circles)
at zero pressure against inverse temperature. 
The dotted line indicates $T_c$. 
The full lines are guides to the eye only.}
\label{fig_amax} 
 \end{center}
\end{figure}

In Fig.~\ref{fig_amax} we show the maximal value of the non-Gaussianity
as function of inverse temperature. In the hot liquid 
$\alpha_2(t)_{\rm max}$ is 0.12 and 0.14 for the two components,
respectively. This corresponds more or less to the vibrational
contribution with very little addition from jump processes. The hot
liquid is, as expected, nearly homogeneous with respect to diffusion.
Cooling down, we observe a marked increase  
of $\alpha_2(t)_{\rm max}$ in the undercooled melt
which accelerates approaching $T_c$. 
At $T_c$ the maximal non-Gaussianity is already 2.5 and 4 for the 
two components, respectively. In the glass,
just about 10\% below $T_{\rm c}$, these values have doubled and 
reach $\alpha_2(t)_{\rm max} \approx 10$
for the smaller component. This
value is still an
underestimate due to ageing effects, see next section.
The non-Gaussianity is strongly pressure dependent. We define
a pressure coefficient as
\begin{equation}
\beta_\alpha(T) = \frac{2}{p_1 - p_2} \cdot
\frac{\alpha_2(T,p_1)-\alpha_2(T,p_2)}{\alpha_2(T,p_1)+\alpha_2(T,p_2)}
\label{eq_alpha_p}
\end{equation}
with $p_1$ and $p_2$ two different applied external pressures.
We find for $T=0.32$ $\beta_\alpha(T) = 0.27$ and $0.35 \epsilon /
\sigma^3$ for the two components,respectively. There is no
drastic effect near $T_c$ ($\beta_\alpha(T) = 0.27$ and $0.35 \epsilon /\sigma^3$).
 
\begin{figure}[ht]
  \begin{center}
 \includegraphics*[bb=30 100 630 530,totalheight=7cm,keepaspectratio]{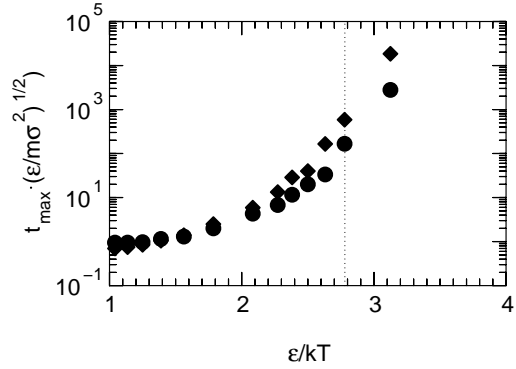}
\caption{Time, $t_{\rm max}$, elapsed till the maximum of the 
non-Gaussianity is reached 
(majority $A$-atoms, diamonds, and minority $B$-atoms, 
circles)
at zero pressure against inverse temperature.  
The dotted line indicates $T_c$.}
\label{fig_tmax}
  \end{center}
\end{figure}

The increase of the maximal value is concomitant with an increase
of the time this value is reached, $t_{\rm max}(T)$ Fig.~\ref{fig_tmax}.
In the hot liquid  $t_{\rm max}(T)$ is of the order of vibrational times
and increases by four orders of magnitude upon cooling to $0.9 T_{\rm c}$.
This slowing 
down reflects the general slowing of the diffusional dynamics as
shown in Fig.~\ref{fig_D_rho}. To check the correspondence between
the diffusion time and $t_{\rm max}$ we calculate the dimensionless
quantity
\begin{equation}
C_{\rm D-NG}(T) = D(T)\rho(T)^{2/3}t_{\rm max}
\label{eq_CDNG}
\end{equation}
where $\rho(T)$ is the density of the system at zero pressure.
In defining $C_{\rm D-NG}(T)$ it is assumed that the non-Gaussianity is
mainly caused by the same atomic motion as diffusion which is indicated
by the rise of $t_{\rm max}(T)$ above the vibrational times upon
undercooling. 
$C_{\rm D-NG}(T)$
should then become independent of temperature.
Indeed the large variation of $t_{\rm max}(T)$ by orders of 
magnitude nearly vanishes. We find,
for both components,  $C_{\rm D-NG}(T=0.56)\approx 0.015$,
at the onset of undercooling, dropping to 
$C_{\rm D-NG}(T=0.32)\approx 0.004$
just below $T_c$. This correlation between $t_{\rm max}(T)$ and
diffusivity is related to the one with the onset of the 
 $\alpha$-relaxation, reported earlier \cite{kob:95b}. 
The drop of $C_{\rm D-NG}(T)$ is probably partially a result of 
the increasing
separation of relaxation and diffusion, the first one being less
sensitive to eventual return events than the long range diffusion..
Another contribution could be a change of the shape  
of the non-Gaussianity versus time curves.

As mentioned in the introduction, there is a general 
$\sqrt(t)$-law governing the increase of $\alpha_2(t)$ 
above its vibrational
value \cite{CMS:00}. From this, one could assume that  
$\alpha_2(t)_{\rm max} / \sqrt{t_{\rm max}(T)}$ should be approximately
constant in the undercooled and glassy regimes. This  holds 
fairly well in the undercooled melt
above $T_{\rm c}$ but breaks down below. Part of this might be due
to ageing but inspection of the actual $\alpha_2(t)$-curves shows that
it is mainly due to an increased flattening near the maximum which
separates  $t_{\rm max}(T)$ from the $\sqrt(t)$-increase.

\section{Ageing}

We have seen that there is a close connection between diffusivity and
dynamic heterogeneity. It is well known that glasses are not in
thermodynamic equilibrium and are subject to ageing. In experiment
one sees upon rapid quenching a drop of the diffusivity on a time
scale of hours \cite{horvath:85}. Ageing, it is generally assumed,
leads to a more
``ideal'' glass. The ``excess volume'' drops which indicates 
annealing of
defects - whatever they are. This poses the question of the
relationship of dynamic heterogeneity and idealness. If the
dynamic heterogeneity is an inherent property of the glassy state
one expects it to increase with ageing as the diffusivity
decreases. 

A simulation over real times of hours is of course
impossible. To circumvent this, ageing effects have been studied by
instantaneous quenches from high temperatures, thus producing large 
effects 
\cite{kob:97b,mussel:98}. We adopted a softer procedure:
samples equilibrated at $T = 0.34 \epsilon /k$ were quenched at a
rate of $Q = 10^{-4}$ to $T = 0.32 \epsilon /k$. 
The quench amounted to about 5\% of $T_{\rm c}$.
The quenched samples 
were then aged and the
diffusion constant was determined 
in subsequent intervals. To determine the diffusion constant at 
time $t$ we calculated the average mean square displacement in the
interval [$t - \delta /2, t + \delta/2$]. The diffusion coefficient was
then obtained by the standard procedure from the long time slope     
in that interval. 

\begin{figure}[ht]
  \begin{center}
 \includegraphics*[bb=30 100 630 530,totalheight=7cm,keepaspectratio]{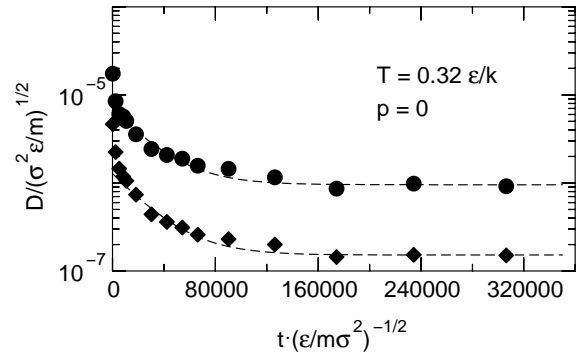}
   \caption{Diffusion coefficient as function of ageing time (majority $A$-atoms, diamonds, and minority $B$-atoms, 
circles). he lines are guides to the eye only. The dashed lines show
the fit with the exponential annealing expression, Eq.~\ref{eq_D_ageing}}
\label{fig_Dageing}
  \end{center}
\end{figure}

Fig.~\ref{fig_Dageing} shows a drop of both diffusivities by about
an order of magnitude over the ageing interval of 80000 time units.
The rapid initial drop of $\log{D(t)}$ is followed by a seemingly  
linear part, dashed lines. 
Such a behavior is consistent with the assumtion that the diffusivity
is the sum of two terms
\begin{equation}
D(t) = D_{\rm inh} + D_{\rm def}c_{\rm def}(t)
\end{equation}
where $D_{\rm inh}$ is a
time independent inherent diffusion coefficient, $D_{\rm def}$
is a defect mediated diffusion constant
and $c_{\rm def}(t)$ is the
defect concentration. If the defects are
slowly annealed, with some rate $\alpha_{\rm def}$ during the 
ageing at constant temperature, we get  
\begin{equation}
D(t) = D_{\rm inh} + D_{\rm def}c_{\rm def}(0)e^{-\alpha_{\rm def} t}.
\label{eq_D_ageing}
\end{equation}
Assuming one type of defect, the same decay constant $\alpha$
should apply to both components whereas the combination
$D_{\rm def}c_{\rm def}(0)$ can vary between them. 
Eq.~\ref{eq_D_ageing} gives, appart from the short time behaviour, 
an excellent fit of the data of 
Fig.~\ref{fig_Dageing}, dashed line. The short time behaviour 
cannot be expected to be reproduced by such a simple model of
independent defects. We find a decay constant 
$\alpha_{\rm def}(T=0.32) = 4.15\cdot 10^{-5}$, and for
the combination $D_{\rm def}c_{\rm def}(0)$ we find values
of $1.45\cdot 10^{-6}$ and $5.84\cdot 10^{-6}$ for the large and
small component respectively. Eq.~\ref{eq_D_ageing} allows us
to extrapolate to the inherent diffusivity which should be reached
after long time aging. These values are shown by the open symbols in 
Fig.~\ref{fig_D_rho}.

Soon after the quench, the first values of the diffusion 
coefficients  
equal within 
10\% the ones before the quench, at the higher temperature. There
are two effects which seem to cancel each other. On the one hand,
due to densification
and temperature reduction, diffusion should slow down. On the other hand,
the quench drives the system further away from equilibrium which

accelerates diffusion processes. It should be noted that diffusion
in this short time interval is not necessarily long range. 

During the ageing the glass is densified. The volume reduction per atom 
is $\Delta \Omega \approx 2\cdot 10^{-3} \Omega$ where $\Omega$
stands for the average atomic volume at that temperature.
There is no direct proportionality with the drop in diffusivity.
By the time the rapid drop finishes ($t \approx 30000$)  
only about 20\% of the excess volume, $\Delta \Omega$, is gone.
The raised diffusivity can, therefore, not be attributed to simple
quasi-vacancies but to more intricate ``defects''.

\begin{figure}[ht]
  \begin{center}
 \includegraphics*[bb=30 100 630 530,totalheight=7cm,keepaspectratio]{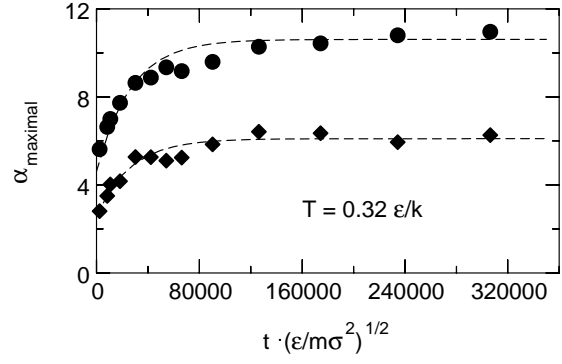}
   \caption{Maximal value of non-Gaussianity as function of ageing time (majority $A$-atoms, diamonds, and minority $B$-atoms, 
circles). The dashed lines show
the fit with the exponential annealing expression, Eq.~\ref{eq_alpha_ageing}.}
\label{fig_aageing}
  \end{center}
\end{figure}

The slowing down of diffusion is accompanied by a
monotonic increase of the
non-Gaussianity, Eq.~\ref{eq_alpha}, by a factor of two over the
ageing period, Fig.~\ref{fig_aageing}
Immediately after the quench the values of 
$\alpha_2(t)_{\rm max}$ are considerably below the ones of the
starting temperature $T = 0.34 \epsilon /k$. This is again consistent
with the above picture that the quench produces some ``defects''
which are annealed during aging. This is described in analogy
to Eq.~\ref{eq_D_ageing} by

\begin{equation}
\alpha_2(t) = \alpha_2^{\rm inh} +\alpha_2 ^{\rm def}c_{\rm def}(0)e^{-\alpha_{\rm def} t}.
\label{eq_alpha_ageing}
\end{equation}

Assuming that the slowing down of diffusion and the increase of
heterogeneity are caused by the same mechanism, we take for the
decay constant the value obtained from the diffusivity. The
resulting fit is shown by the dashed lines in Fig.~\ref{fig_aageing}.
Considering the obvious scatter the fit is again excellent.
The resulting values are for the inherent value
$\alpha_2^{\rm inh} = 6.2$ and $10.0$ and for the defect part
$\alpha_2 ^{\rm def}c_{\rm def}(0) = -4.0$ and $-5.1$ for the
two components, respectively. 

The close connection between collective jumps and heterogeneity
gives a hint of the possible nature of what we loosely call defects.
Defect here does not mean a simple structure such a vacancy but more likely a
local center of strain. These local strains can be relaxed by
a less collective motion than the one in the relaxed glass. This was
found earlier by tracer experiments experiments which investigated 
the effect of ageing
on the isotope effect \cite{ratzke:92}. This additional motion
apparently is more homogeneous than the one inherent to the
relaxed glass. At the early stages or after a large rapid quench,
collectivity might be fully destroyed for some jumps.
In a simulation of a BLJ glass at constant density
single particle jumps have been observed after a rapid quench
\cite{vollmayr:03}.

Finally,
checking the correlation, Eq.~\ref{eq_CDNG} we find
a drop by a factor of 2-3 during ageing, similar to the one found
in the temperature dependence. We interpret this as an indication
that the excess diffusivity produced the quench shows less decoupling
between diffusion and relaxation than the inherent one.

\section{Summary}

We investigate by molecular dynamics the relation between diffusivity
and dynamic heterogeneity. We use a binary Lennard-Jones like system
as simple model of a metallic glass, respectively melt. This model
system has been extensively studied, mostly at constant volume. In this
study the volume is varied to have zero pressure conditions. The 
diffusion coefficients in the melt are in accordance with mode
coupling theory with a single critical temperature $T_{\rm c}$.
In the undercooled melt and, even more pronounced in the glass,
diffusion is strongly heterogeneous over long times. The 
distribution of atomic displacements deviates from the Gaussian
distribution found for random walks. 

The non-Gaussianity parameter increases, from its small starting
value due to vibrations, initially with a $\sqrt{t}$-law before
long range diffusion finally restores homogeneity. The time of the
maximum non-Gaussianity is given approximately by the diffusion 
time. However, 
the correlation factor between the two times 
decreases systematically with reducing temperature. This may 
indicate a decoupling between relaxations and long range diffusion
as has been observed in studies of viscosity versus diffusivity.
Below $T_g$ the $\sqrt{t}$-increase does no longer determine
the maximal value of the non-Gaussianity since the maximum
rapidly flattens.

After a quench below $T_c$ we find a drop of the diffusion 
coefficients by
an order of magnitude with ageing at constant temperature. 
This can be
explained in terms of defects which are produced in the quench and
are subsequently annealed. These defects lead to faster diffusion
and lower heterogeneity. In this simple system, defect probably means 
some center of
local strain which allows for a less collective motion. 
In a more complicated system ``defects'' could, e.g., be different molecular
conformations \cite{chelli:03}. This general
picture is supported by the fact that in the present binary system the 
aging of both, diffusivities and heterogeneity, of both components
can be described by a single decay constant. This can be used
to extrapolate to the inherent diffusion coefficients of the 
ideal glass at the given temperature
At the temperatures accessible to simulation this ideal glass
would actually still be an undercooled liquid, alas with
$T<T_{\rm c}$. 
The heterogeneity increases with ageing.
We conclude that it is an inherent property of the relaxed glass
which is suppressed by defects.


\clearpage
\begin{list}{}{\leftmargin 2cm \labelwidth 1.5cm \labelsep 0.5cm}

\item[\bf Fig. 1] 
Diffusion constants (majority $A$-atoms, diamonds, and minority $B$-atoms, 
circles) and density (triangles)
at zero pressure against inverse temperature (all in reduced units).
The open symbols indicate the extrapolated inherent diffusion
coefficients according to Eq.~\ref{eq_D_ageing}
The dashed lines
show the fit with the mode coupling expression for atoms $A$ and $B$,respectively.
\item[\bf Fig. 2]
Maximum value of the non-Gaussianity  
(majority $A$-atoms, diamonds, and minority $B$-atoms, 
circles)
at zero pressure against inverse temperature.
The dotted line indicates $T_c$. 
The full lines are guides to the eye only.
\item[\bf Fig. 3] 
Time, $t_{\rm max}$, elapsed till the maximum of the non-Gaussianity 
is reached 
(majority $A$-atoms, diamonds, and minority $B$-atoms, 
circles)
at zero pressure against inverse temperature.
The dotted line indicates $T_c$.
\item[\bf Fig. 4]
Diffusion coefficient as function of ageing time (majority $A$-atoms, diamonds, and minority $B$-atoms, 
circles). The dashed lines show
the fit with the exponential annealing expression, Eq.~\ref{eq_D_ageing}.
\item[\bf Fig. 5]
Maximal value of non-Gaussianity as function of ageing time (majority $A$-atoms, diamonds, and minority $B$-atoms, 
circles). The dashed lines show
the fit with the exponential annealing expression, Eq.~\ref{eq_alpha_ageing}.
\end{list}
\clearpage

\clearpage

\end{document}